\documentclass[twocolumn]{aastex6}

\begin{document}
\title{Descattering of Giant Pulses in PSR B1957+20}
\author{Robert Main$^{1,2,3}$, Marten van Kerkwijk$^{1}$, Ue-Li Pen$^{2,3,4,5}$, Nikhil Mahajan$^{1}$, Keith Vanderlinde$^{1,3}$}
\affil{
$^1$Department of Astronomy and Astrophysics, University of Toronto, 50 St. George Street, Toronto, ON M5S 3H4, Canada \\
$^2$Canadian Institute for Theoretical Astrophysics, University of Toronto, 60 St. George Street, Toronto, ON M5S 3H8, Canada \\
$^3$Dunlap Institute for Astronomy and Astrophysics, University of Toronto, 50 St George Street, Toronto, ON M5S 3H4, Canada \\
$^4$Perimeter Institute for Theoretical Physics, 31 Caroline Street North, Waterloo, ON N2L 2Y5, Canada \\
$^5$Canadian Institute for Advanced Research, 180 Dundas St West, Toronto, ON M5G 1Z8, Canada \\
}

\begin{abstract}
The interstellar medium scatters radio waves which causes pulsars to scintillate.  For intrinsically short bursts of emission, the observed signal should be a direct measurement of the impulse response function.  We show that this is indeed the case for giant pulses from \object{PSR B1957+20}: from baseband observations at 327\,MHz, we demonstrate that the observed voltages of a bright pulse allow one to coherently descatter nearby ones.  We find that while the scattering timescale is $12.3\,\mu$s, the power in the descattered pulses is concentrated within a span almost two orders of magnitude shorter, of $\lesssim\!200\,$ns.  This sets an upper limit to the intrinsic duration of the giant pulses.  We verify that the response inferred from the giant pulses is consistent with the scintillation pattern obtained by folding the regular pulsed emission, and that it decorrelates on the same timescale, of~$84\,$s. In principle, with large sets of giant pulses, it should be possible to constrain the structure of the scattering screen much more directly than with other current techniques, such as holography on the dynamic spectrum and cyclic spectroscopy.
\end{abstract}

\section{Introduction and Background}
\label{sec:introduction}
\noindent


When observed at relatively low frequency, pulsars scintillate, showing an interference pattern in frequency and time that arises from multi-path propagation through the interstellar medium.  The interference patterns often have clear structure, showing a quadratic dependence of the delay of scattering points as a function of their fringe rate (leading to parabolic arcs in the secondary spectra), which is most readily understood if the scattering is dominated by localized points on a strongly anisotropic screen \citep{stinebring+01, walker+04, cordes+06}. This picture was confirmed dramatically by \citealt{brisken+10}, who used very long baseline interferometry to show that the scattering screen of \object{PSR B0834+06} appeared on the sky as a collection of points along a single, linear structure.

The above not only provides surprising information about the nature of the interstellar medium, but also offers a remarkable opportunity to study pulsars: given sufficient understanding of the locations of the scattering points, one can use them as an interferometer, which, with baselines of tens of AU, has sub-microarcsecond angular resolution, thus allowing precision astrometry.  Indeed, \cite{pen+14} et al.\ used the scintillation for \object{PSR B0834+06} to show that the location of the radio emission shifted by a few 10s of km as a function of spin phase.

The scintillation properties of pulsars are typically inferred from their dynamic spectrum, i.e., the intensity of a pulsar's folded emission as a function of frequency and time. This yields direct information on the amplitudes of the interstellar impulse response function, but to retrieve its phase one has to rely on holographic techniques \citep{walker+08, pen+14}. A promising alternative is to use cyclic spectra of pulsars \citep{demorest11}, which retain part of the phase information.  So far, however, both methods have been shown to work only in specific cases.


In principle, the interstellar response could be measured directly if an object emitted bursts of emission that lasted much shorter than the scattering time.  Some pulsars oblige by emitting suitable short ``giant pulses.'' One of these is the ``black widow'' pulsar \object{PSR B1957+20} \citep{knight+06}.  In this paper, we show that its giant pulses indeed allow one to measure the interstellar response directly.

\section{Giant Pulses}

We recorded 9.5\,hr of P-band data of PSR B1957+20 at the Arecibo Observatory, as part of a European VLBI network program (GP~052).  The data were taken in four daily 2.4\,hr sessions on 2014 June 13--16, recording dual circular polarizations of four contiguous 16\,MHz wide bands spanning 311.25 to 375.25\,MHz (in 2-bit Mark~4 format). We exclude the fourth band from our analysis, as its signal was almost fully filtered out by the receiver, as well as the June 15 data, as these cover the eclipse of the pulsar by the wind of its companion \citep{fruchter+88}, which hinders our analysis. We are thus left with 7.2\,hr of data covering 311.25--359.25\,MHz.

No flux calibrators were observed, so we convert to flux based on a nominal system temperature of 120\,K and gain of $10{\rm\,K/Jy}$ for the 327\,MHz receiver.\footnote{http://www.naic.edu/~astro/RXstatus/327/327greg.shtml} With these values, the folded profile yields an average flux of $37\,$mJy, consistent with the $38\pm3$\,mJy found by \citet{fruchter+92}.

We searched for giant pulses in the de-dispersed time streams of both polarizations (using a dispersion measure of $29.1162{\rm\,pc\,cm^{-3}}$, tweaked using the folded profile), by binning the power in the whole 48\,MHz band to $16\,\mu$s resolution and flagging peaks above $12\sigma$ ($\sim\!3\,$Jy).  We found $247$ and $313$ pulses in left and right circular polarization, respectively, with $102$ of these in common.\footnote{At lower thresholds, many more true giant pulses are present, but separation from the bright tail of the ``regular'' pulses becomes more difficult, and these fainter pulses are less useful for our purposes here.}

\begin{figure*}
\includegraphics[width=1.0\textwidth]{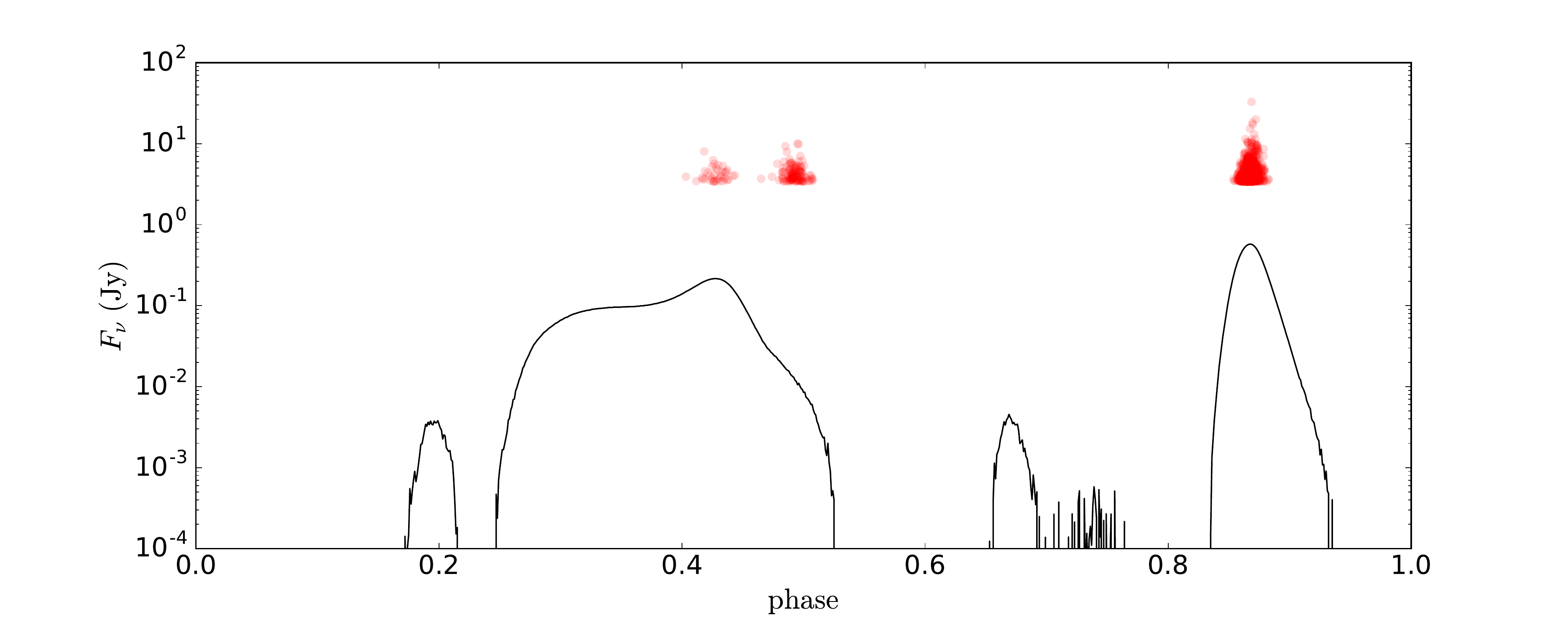}
\caption{The folded profile (solid curve) and the giant pulses (points) as a function of pulse phase. The fluxes of the giant pulses are as measured in the $16\,\mu$s-binned timestream used to find them.}
\label{fig:phases}
\end{figure*}

In Figure~\ref{fig:phases}, we show where the giant pulses arrive relative to the average profile.  One sees three clusters, with the first, containing most pulses, coincident with the main pulse, the second coincident with the peak of the interpulse, and the third on the (relatively weak) tail of the interpulse. This distribution is different from what is seen in \object{PSR B1937+21}, where the giant pulses are predominantly found at the trailing edges of the pulse components \citep{cognard+96, soglasnov+04}.


\begin{figure}
\includegraphics[width=1.0\columnwidth]{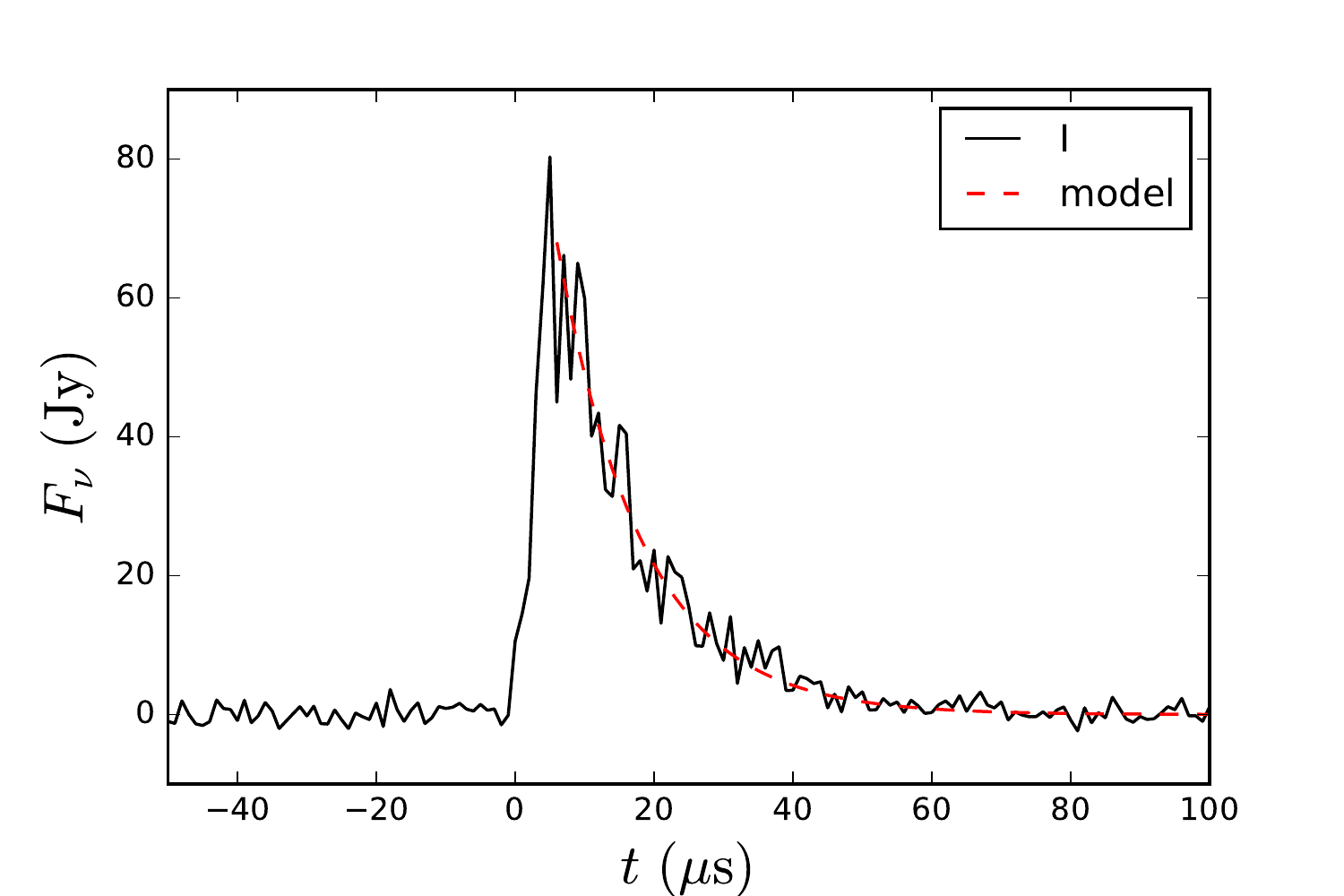}\\
\includegraphics[width=1.0\columnwidth]{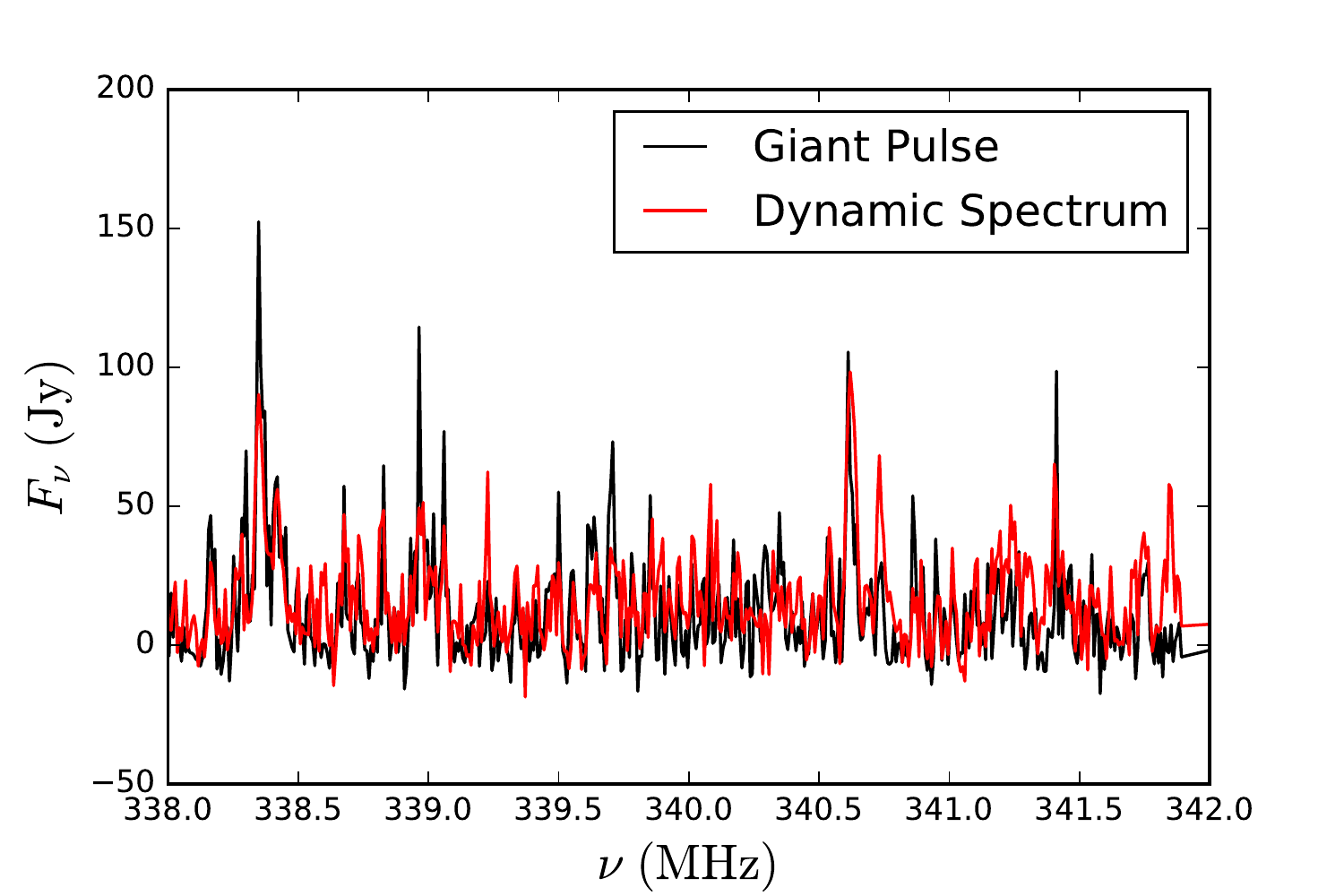}
\caption{{\it Top:\/} Profile of the brightest giant pulse in $1\,\mu$s bins, with an exponential fit with a timescale of $12.3\,\mu$s overlaid on its tail. {\it Bottom:\/} A segment of the power as a function of frequency for the brightest giant pulse compared with that for the regular emission (determined using 30\,s around the pulse, and scaled to match its flux). The bins are 8\,kHz wide, the highest resolution possible given the width of the main pulse in the folded profile.}
\label{fig:brightest}
\end{figure}

The profiles of the giant pulses show a sharp rise and a long tail, suggesting that the pulses are intrinsically short.  The tails are well fit by an exponential, with a timescale of $12.3\,\mu$s (see Fig.~\ref{fig:brightest} lower panel). The pulses are close to $100\%$ polarized, as expected for intrinsically short, single mode emission. The giant pulses show no preferred polarization direction, and can be strongly linearly or circularly polarized in either direction, in contrast to the folded profile, which is close to unpolarized \citep{fruchter+90}.  This allows us to measure a rotation measure for the source; fitting a rotation measure to Stokes Q of the brightest pulse gives $ \text{RM} = 46.3 \pm 0.7 \text{ rad / m}^{2} $.

\section{Scintillation and Scattering}
\label{section:scattering}

If giant pulses and the regular pulsar emission are both affected in the same way by propagation through the interstellar medium, an immediate expectation is that their frequency power spectra should show similar structure, and that this structure should vary on the same timescale.  Comparing the power spectra for the brightest giant pulse with that of the regular emission near it, we indeed find that the spectra are very similar (see Fig.~\ref{fig:brightest}).

A stronger expectation is that the impulse response is the same, i.e., for giant pulses that happen close in time, not just the amplitudes of their spectra should match, but also their phases.  We first verify this is the case for our closest pair, and then show it is possible to use a giant pulse as a direct measurement of the response function.

\begin{figure}
\includegraphics[width=1.0\columnwidth]{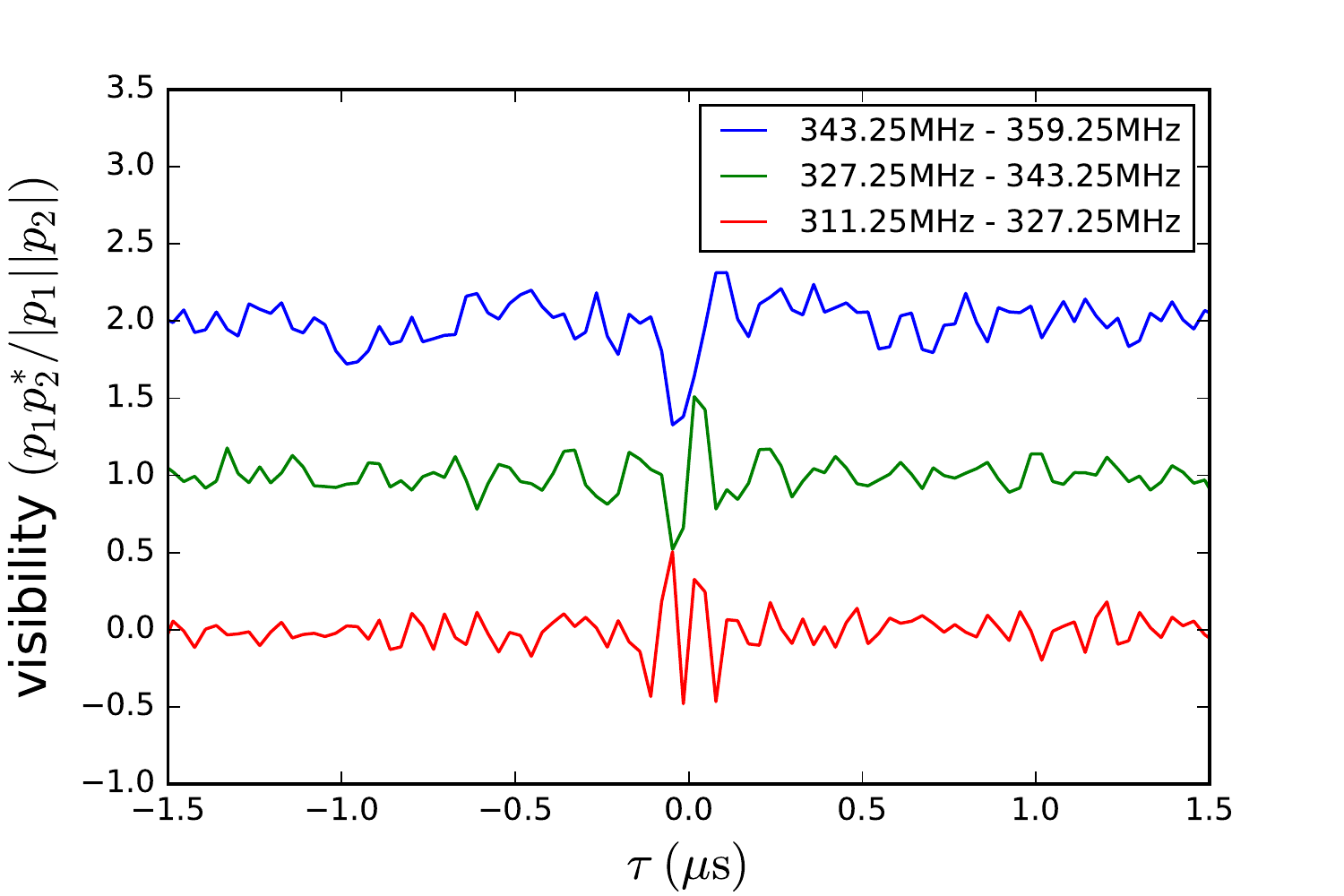} 
\caption{Correlations between the voltage streams of a bright pulse pair separated by 1.92\,s, for the three 16\,MHz bands we observed in. This is equivalent to computing the visibility between the two pulses, as if they were a single pulse observed at two telescopes. The power is spread over $\sim\!200\,$ns, either because of small differences in the interstellar response, or, more likely, because of intrinsic differences in the two pulses. In either case, this spread sets an upper bound on the intrinsic duration of both pulses.}
\label{fig:vis}
\end{figure}

\subsection{Giant Pulse Intrinsic Width}

We directly compare the voltage timestreams of the two closest giant pulses, which are separated by 1.922\,s.  Both are strongly right-hand polarized, with integrated signal-to-noise of 46 and 18, respectively.  We compute visibilities between the two pulses as a function of time lag in each of the three 16\,MHz, right-hand circular bands.  In Figure~\ref{fig:vis}, we show the visibility divided by the geometric mean of the auto-correlations, i.e., a measure of the correlation strength of the voltage streams.  We find strong correlation within a narrow envelope, which proves that the response of the interstellar medium is indeed close to identical for these two pulses, and that their intrinsic durations are very short, $\lesssim\!200\,$ns.

\subsection{The Interstellar Response Function}
\label{section:descatter}

The observed data are the convolution of the intrinsic electric field $E_{\rm int}$ with the impulse response function $g$ of the interstellar medium, i.e., $E_{\rm obs}(t) = (E_{\rm int} * g)(t)$.  To the extent that giant pulses approximate impulsive intrinsic emission, it is thus possible to use them to measure the response, with which it should then be possible to undo the effects of scattering and scintillation, at least within the decorrelation time $t_{\rm corr}$ on which the response changes.

To verify this, we attempt to measure how well we can ``descatter'' pairs of giant pulses with each other.  We do our analysis in Fourier space, where the pulses can be described as 
\begin{eqnarray}
  \tilde{E}_{\rm obs}(\nu) &=& {\cal F}(E_{\rm obs}(t))\nonumber\\
  &=& {\cal F}((E_{\rm int} * g)(t)) = \tilde{E}_{\rm int}(\nu) \tilde{g}(\nu)
\end{eqnarray}
If the intrinsic emission is a delta function at $t=0$, one would have a Fourier spectrum with constant amplitude and zero phase, but because of the interstellar response, the signal gets mixed between frequency channels, causing amplitudes to change and phases to rotate (but with total power conserved).

In principle, the above suggests we could descatter a giant pulse by dividing by a suitably normalised reference pulse.  In practice, this is rather noisy, as low-amplitude and thus noisy channels get upweighted and high-amplitude and thus well-measured ones downweighted.  Instead, therefore, we first normalise the reference pulse at all frequencies, i.e., $\tilde{E}_{\rm obs}^{\rm ref}(\nu) / |\tilde{E}_{\rm obs}^{\rm ref}(\nu)|$.  Dividing the trial pulse by this gives,
\begin{equation}
  \frac{\tilde{E}_{\rm obs}^{\rm trial}}
       {\tilde{E}_{\rm obs}^{\rm ref} / |\tilde{E}_{\rm obs}^{\rm ref}|}
= \left(\frac{\tilde{E}_{\rm int}^{\rm trial}}
             {\tilde{E}_{\rm int}^{\rm ref} / |\tilde{E}_{\rm int}^{\rm ref}|}
  \right)
  \left(\frac{\tilde{g}^{\rm trial}}
             {\tilde{g}^{\rm ref}/|\tilde{g}^{\rm ref}|}
  \right),
\label{eq:descatter}
\end{equation}
where we dropped the dependence on $\nu$ for brevity.

If the pulses have the same response function and the reference pulse is truly impulsive, this reduces to $\tilde{E}_{\rm int}^{\rm trial}|\tilde{g}|$, i.e., one would recover the intrinsic spectrum of the trial pulse multiplied by the amplitudes of the response function.  Since the phases are corrected but the amplitudes are not, for an intrinsically short pulse, an inverse Fourier transform should yield a timestream with a pulse which is similarly short but which has reduced amplitude (as discussed quantitatively below).  More generally, since we do not have perfect arrival times, there will be an uncertainty in the time offset (equivalent to a phase gradient in the spectrum), and, since the intrinsic pulses are not true delta functions, the power will only be concentrated within the intrinsic width of the emission.  

We apply our method first using our brightest giant pulse as a reference, and its five closest neighbours as trial pulses. Our brightest pulse is strongest in right-circular polarization, with an integrated signal-to-noise of 132. We make Fourier transforms for $32\,\mu$s segments for all pulses, covering the majority of the scattering tail (this corresponds to 1024 real-valued samples and thus 512 channels in each of the three 16\,MHz bands). We then descatter and inverse transform as above, and bin and sum the power of the descattered timestreams in 250\,ns bins to account for possible intrinsic widths (see Fig.~\ref{fig:vis}).

\begin{figure*}
\centering
\includegraphics[trim=0 0.5cm 0.8cm 0.2cm, width=0.75\hsize, clip]{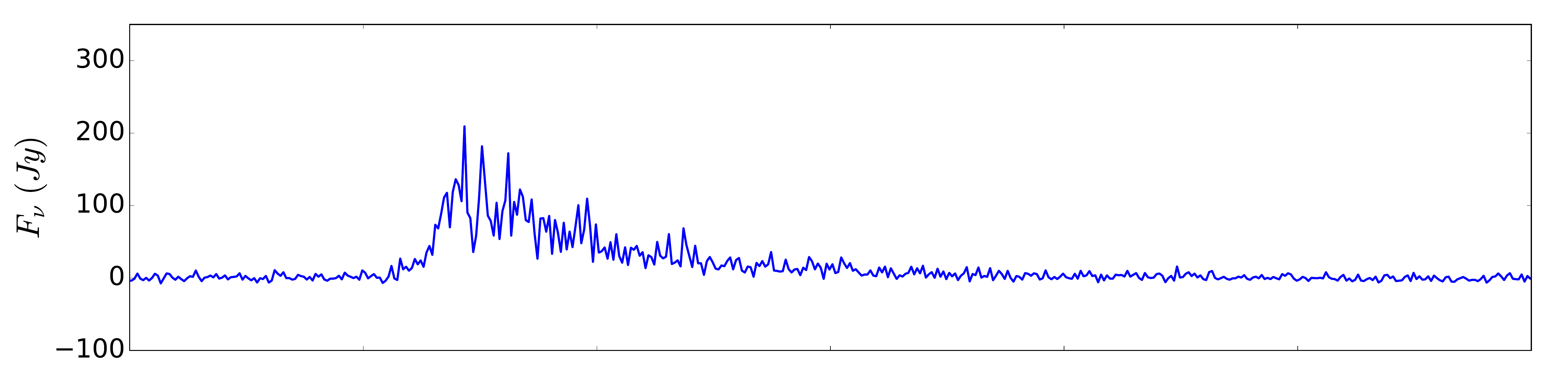}\\
\includegraphics[trim=0 0.5cm 0 0.2cm, width=0.75\hsize, clip]{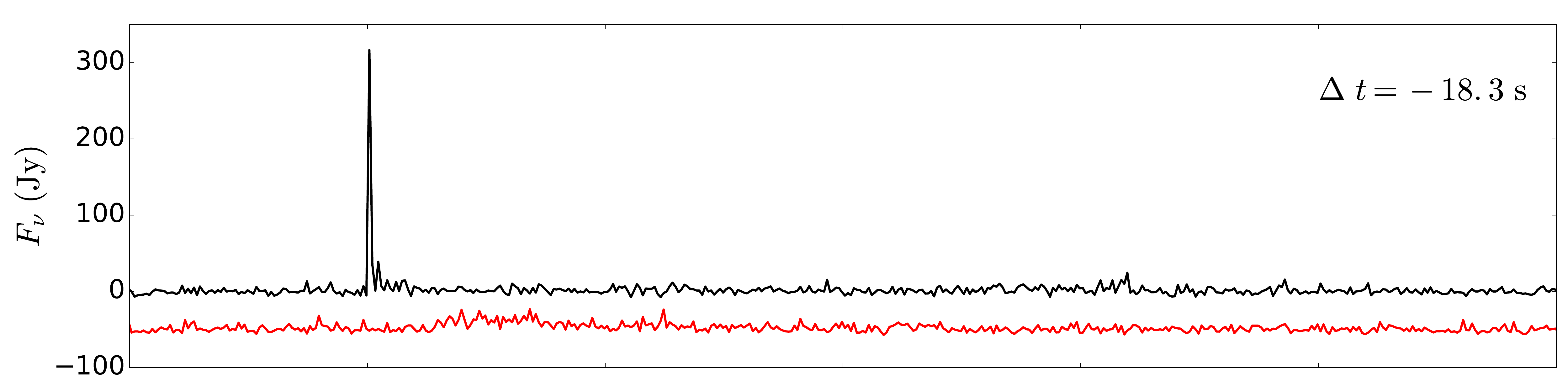}\\
\includegraphics[trim=0 0.5cm 0 0.2cm, width=0.75\hsize, clip]{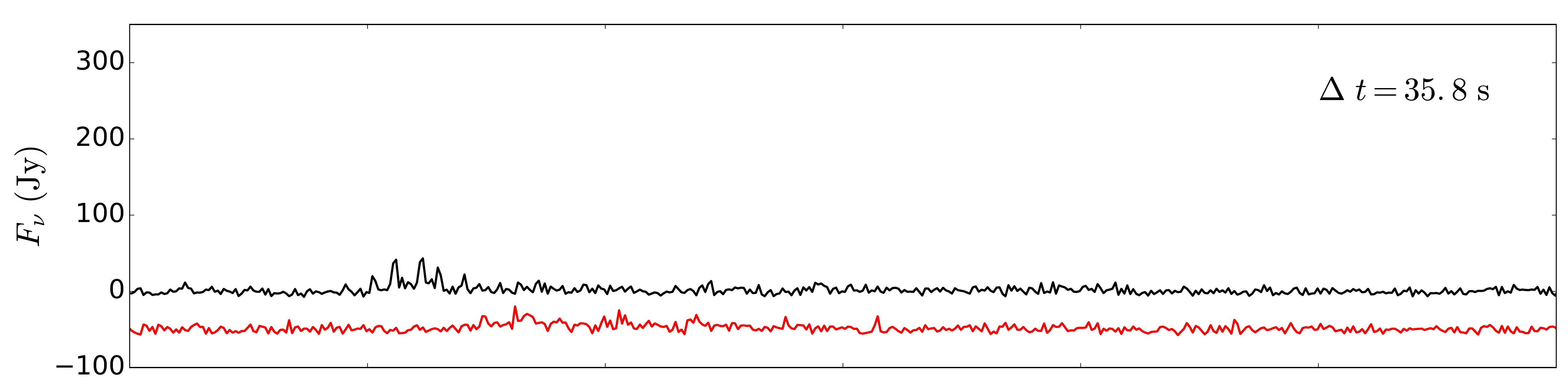}\\
\includegraphics[trim=0 0.5cm 0 0.2cm, width=0.75\hsize, clip]{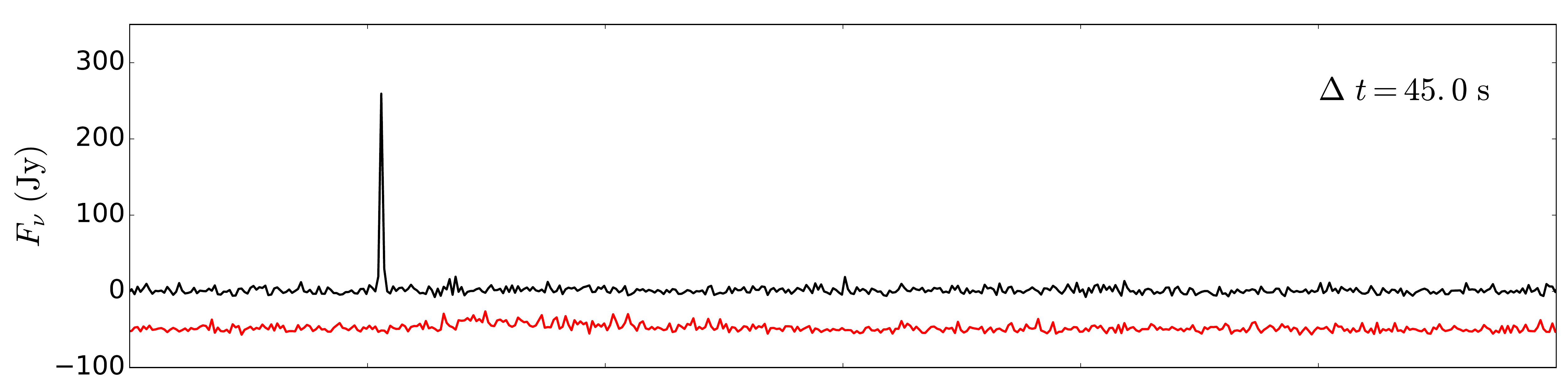}\\
\includegraphics[trim=0 0.5cm 0 0.2cm, width=0.75\hsize, clip]{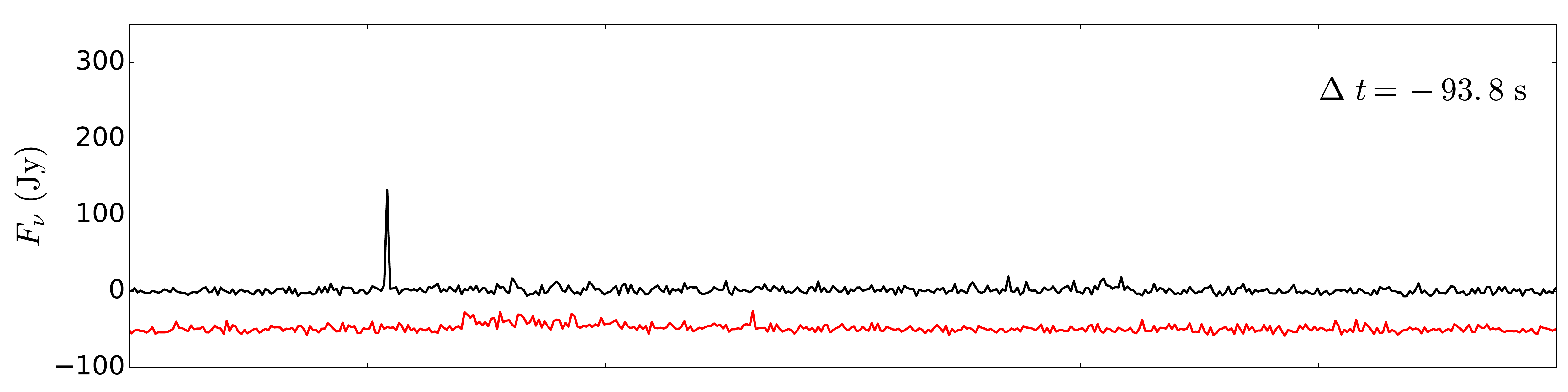}\\
\includegraphics[trim=0 0 0.8cm 0.2cm, width=0.75\hsize, clip]{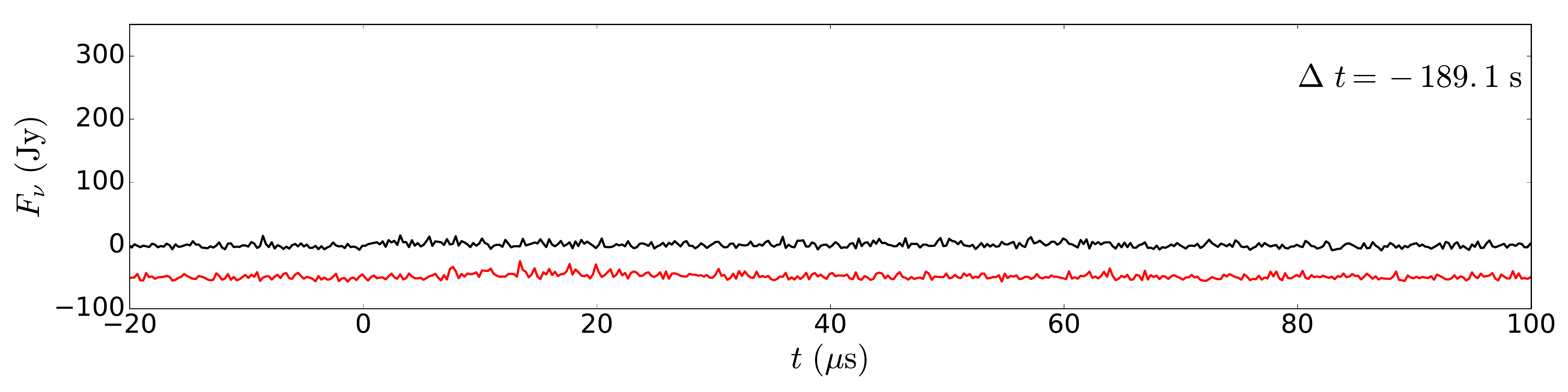}
\caption{Descattering giant pulses. {\em Top:\/} The profile of the brightest giant pulse in our sample, used to descatter its neighbouring pulses. {\em Lower panels:\/} Nearby giant pulses, ordered by time separation, showing both the profiles of the observed, scatter-broadened pulse (red curve, lowered by 50 Jy), and the descatted pulse (black curve). For all pulses, the signal is binned to 250\,ns. One sees that the descattering works less well at larger time differences, for which the interstellar response starts to change significantly. The outlier at $\Delta t = 35.8\,$s appears to be formed of a series of bursts, in contrast to the other pulses which appear to be well described by delta functions in this time binning.}
\label{fig:descatter}
\end{figure*}

The results are shown in Figure \ref{fig:descatter}.  For the closest pair, one sees that much of the power of the descattered pulse is contained within a single 250\,ns bin, and that the peak intensity of the descattered pulse is more than 10 times stronger than that of the observed, scatter-broadened one.  For the next closest pulse, the procedure does not seem to work well, with the descattered pulse having multiple peaks.  This is likely intrinsic, since for the further pulses, the descattering does work, though with decreasing efficiency, presumably because the response function becomes increasingly different.

\subsection{Decorrelation Time}

The above suggests it should be possible to use the extent to which giant pulses can descatter each other to determine the timescale on which the response function changes. For this purpose, we increase our sample, by selecting as reference pulses all those for which phases can be measured in the Fourier spectrum, i.e., with $\gtrsim\!1\sigma$ per voltage sample, corresponding to an integrated signal-to-noise of~$\>\!40$ in the power. This results in 6 and 3 pulses in the right and left circular time streams, respectively (with 1 in common). For each of these, we descatter all pulses within 5 minutes to either side, matching the polarizations. We then sum the power of the descattered timestreams in $1\,\mu$s bins (which should account for any reasonable intrinsic widths), and measure the fraction of the total power which is descattered into a single peak to quantify how well the pulses descattered each other.

With our brightest pulse, we find that we can recover up to $70\%$ of the total flux in the descattered peak.  For the other, fainter pulses, however, at most half of the power is descattered, as their more limited strength allows only an imperfect measure of the response function. To quantify the imperfection, we simulate giant pulses, using delta functions at a range of strengths convolved with a response simulated by drawing from a normal distribution with zero mean and variance varying as an exponential with the observed $12.3\,\mu$s decay time (and normalized to have unity integrated power). We then create pairs with the same response function but different amplitudes and independent noise, and run these through our analysis routines.

As expected, the strength of the reference pulse determines the average power recovered, while the strength of the (fainter) trial pulse dominates the scatter between different realisations. For our set of reference pulses, we find the simulated recovered fraction ranges from 0.4 to~0.7. From Eq.~\ref{eq:descatter}, one sees that in the high signal-to-noise limit one should be dominated by the extent to which $\langle|\tilde{g}|\rangle^2$ is less than one. Since our simulations assume $2\tilde{g}^2$ is distributed as a $\chi^2$ distribution with 2 degrees of freedom, one expects $\langle|\tilde{g}|\rangle^2=\pi/4$, consistent with our results.  For lower signal-to-noise, we find that we can also reproduce our simulated fractions by integrating numerically over probability distributions for both $\tilde{g}$ and the noise (unfortunately, we could not find a closed-form expression).

\begin{figure}
\includegraphics[width=1.0\columnwidth]{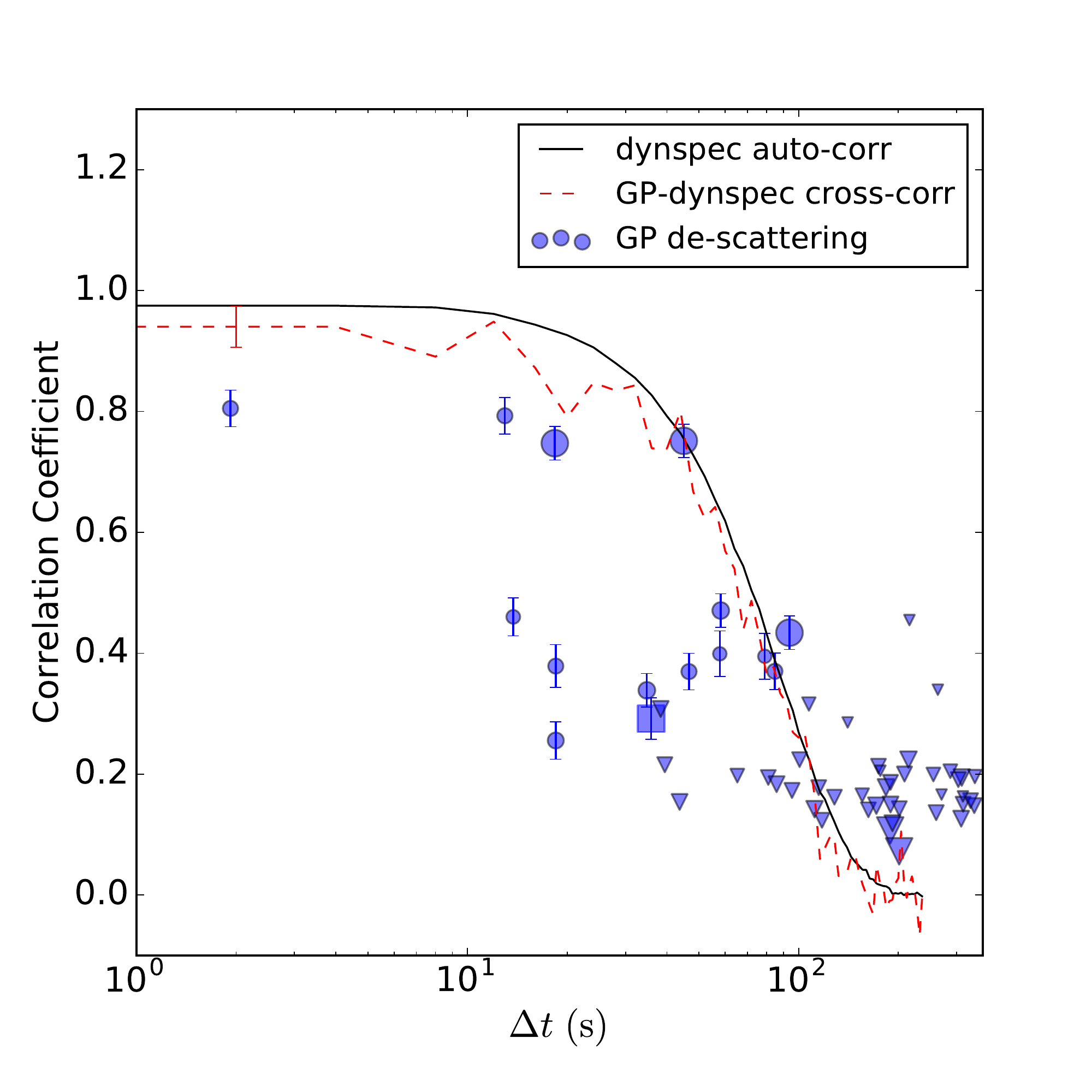} 
\caption{Degree of correlation of pulsar emission as a function of time, determined using three methods. The first is the standard auto-correlation of the dynamic spectrum (solid black curve; measurement errors smaller than the width of the curve, but larger systematic uncertainties; see Sect.~\ref{section:dynspec}).  The second is a cross-correlation of the spectrum of the brightest giant pulse with the dynamic spectrum (dashed red curve; uncertainty estimated from scatter at large $\tau$).  The third uses the fraction of the power of a giant pulse that is descattered by using a bright neighbouring giant pulse as a reference measurements of the impulse response (blue points with error bars, with size scaled to reflect the strength of the reference pulse).  Here, the power has been corrected for the expected loss given the signal-to-noise ratio of the reference pulse (see Sect.~\ref{section:descatter}), and the error bars reflect the uncertainty due to noise in the pulse being descattered.  Points less than $3\sigma$ above the noise inferred from non-pulse parts of the descattered timestream are shown as upper limits. The square point corresponds to the multi-peaked giant pulse shown in Figure 4.}
\label{fig:timescale}
\end{figure}

In Figure \ref{fig:timescale}, we show the fraction of the power that is descattered against time separation for each pulse pair, corrected for the above loss of power (and with error bars reflecting the expected $1\sigma$ scatter). One sees that at large separation, $\gtrsim\!100\,$s, the descattering never recovers much power, while at shorter separation it does, though unequally so.  Inspection of the low points around $\Delta t=20\,$s shows that in those cases the descattering does not lead to a single strong pulse (as for the $\Delta t=35.8\,$s pulse in Fig.~\ref{fig:descatter}).  Likely, this reflects intrinsic pulse structure, with it appearing likely from inspection that some of the reference pulses contain intrinsic structure.

\label{section:dynspec}

We can compare our decorrelation timescale with that derived using the more traditional way, from the autocorrelation of the dynamic spectrum,
\begin{equation}
R(\tau) = \frac{\langle(I(t)-\mu) (I(t+\tau)-\mu)\rangle}{\sigma^{2}},
\end{equation}
where $\mu=\langle I(t)\rangle$ and $\sigma^2=\langle(I(t)-\mu)^2\rangle$.

We apply this to a dynamic spectrum created for the 9 minutes of data surrounding our brightest giant pulse.  We use bins of 4\,s, 8\,kHz, and 1/32 in phase (where the frequency and phase resolution are set to barely resolve the main pulse; at 8\,kHz, we also barely resolve the frequency structure due to scintillation).  We define two off gates, with one subtracted from the folded profile to give a pulsed flux, and the other used as an independent measure of the noise in the dynamic spectrum.  We then compute the auto-correlation of the dynamic spectrum, subtracting the auto-correlation of the noise, and averaging over the central 14\,MHz of each band (to avoid the parts most affected by bandpass variations).

The result is shown in Figure~\ref{fig:timescale}.  One sees that at short times the correlation is very good (it approaches 0.98 rather than 1, likely because we ignore the frequency dependence of the noise), and then it decreases smoothly. Taking the decorrelation time as the lag where the correlation drops by $1/e$, we find $t_{\rm corr}=84\,$s.

In Section~\ref{section:scattering}, we already showed that the spectrum of brightest pulse was similar to that of the regular emission.  With the dynamic spectrum, we can verify this quantitatively, and also check the dependence on lag. For this purpose, we Fourier transform our brightest giant pulse to the same channelization as the dynamic spectrum (where 8\,kHz corresponds to $125\,\mu$s in time, i.e., it covers the full width of the scattering tail; see Fig.~\ref{fig:brightest}).  We then correlate it against the dynamic spectrum for a range of delays.  

At low delay, we find that the giant pulse correlates very strongly with the dynamic spectrum, at $92\pm5\%$ (see Fig.~\ref{fig:timescale}). This strong correlation independently suggests a short intrinsic duration of the giant pulse, as intrinsic structure on timescales comparable to the scattering time will lead to differences in the spectra (as seen for the scintillation pattern of the Crab's giant pulses, \citealt{cordes+04}).  At longer delays, the correlation drops, and we find $t_{\rm corr}=84\,$s, the same value obtained from the auto-correlation of the dynamic spectrum.

Comparing with the points from descattering giant pulses with each other, one finds the curves derived using the power spectra form a rough upper envelope, suggesting the decorrelation time is the same.  This is not a trivial comparison, since the auto-correlations take into account the amplitudes of the impulse response function only, while our descattering only uses the phase information.

\section{Ramifications}

We have shown that giant pulses in \object{PSR~B1957+20} can be used as a direct probe of the impulse response function of the interstellar medium: they allow one to descatter other giant pulses. An immediate result is that we can constrain the typical intrinsic duration of giant pulses of \object{PSR~B1957+20} to be very short, $\lesssim\!200\,$ns.

Having a direct measure of the response also should allow one to verify (and inform) the response inferred from the dynamic spectrum (using holographic techniques; \citealt{walker+08,brisken+10}) or from cyclic spectra \citep{demorest11,walker+13}, thus putting those techniques on firmer footing.

Longer term, one might hope to use all giant pulses -- perhaps aided by the dynamic and cyclic spectra -- to measure the evolution of the response as a function of time and frequency.  If the scattering towards PSR~B1957+20 is dominated by a single, highly anisotropic screen, as was found for \object{PSR~B0834+06} \citep{brisken+10}, then this should allow one to determine amplitudes and phases of individual scattering points directly.  Those, in turn, might allow one to resolve the pulsar's orbit on the sky, as has been done for the pulse emission with spin phase for \object{PSR~B0834+06} \citep{pen+14}.

Unfortunately, only few other pulsars show giant pulses. Among those, we are most excited to apply our technique to the \object{Crab pulsar}, since for that source we expect that the main scattering screen, which is in the Crab nebula, will resolve the light cylinder, thus opening up the possibility to determine empirically where the giant pulses originate.

\acknowledgements
We thank Aleksandar Rachkov for an initial calculation of the dynamic spectra, and Franz Kirsten, I-Sheng Yang, and Dana Simard for helpful discussions.  We made use of NASA'a Astrophysics Data System and SOSCIP Consortium’s Blue Gene/Q computing platform.

\facility{EVN:Arecibo:327-MHz Gregorian}

\software{Astropy \citep{astropy13}}

\bibliographystyle{apj}

\end{document}